\documentclass[12pt]{article}

\newsavebox{\ns}
\newsavebox{\dbrane}
\newsavebox{\dbshort}

\usepackage{epsfig}
\usepackage{amssymb}

\def\appendix{{\newpage\section*{Appendix}}\let\appendix\section%
        {\setcounter{section}{0}
        \gdef\thesection{\Alph{section}}}\section}

\def\be{\begin{eqnarray}}
\def\ee{\end{eqnarray}}

\newcommand{\nn}{\nonumber}

\newcommand{\ft}[2]{{\textstyle\frac{#1}{#2}}}
\newcommand{\eqn}[1]{(\ref{#1})}

\def\Dslash{\,\,{\raise.15ex\hbox{/}\mkern-12mu D}}
\def\Dbarslash{\,\,{\raise.15ex\hbox{/}\mkern-12mu {\bar D}}}
\def\delslash{\,\,{\raise.15ex\hbox{/}\mkern-9mu \partial}}
\def\delbarslash{\,\,{\raise.15ex\hbox{/}\mkern-9mu {\bar\partial}}}
\def\pslash{\,\,{\raise.15ex\hbox{/}\mkern-9mu p}}
\def\calDslash{\,\,{\raise.15ex\hbox{/}\mkern-12mu {\cal D}}}

\begin{document}
\pagestyle{plain}
\setcounter{page}{1}
\newcounter{bean}
\baselineskip16pt

\begin{titlepage}

\begin{center}
\today
\hfill hep-th/0210010\\
\hfill MIT-CTP-3311 \\

\vskip 1.5 cm
{\large \bf The Moduli Space of Noncommutative Vortices}
\vskip 1 cm 
{David Tong}\\
\vskip 1cm
{\sl Center for Theoretical Physics, 
Massachusetts Institute of Technology, \\ Cambridge, MA 02139, U.S.A.\\}

\end{center}

\vskip 0.5 cm
\begin{abstract}
The abelian Higgs model on the noncommutative plane admits both BPS vortices and non-BPS 
fluxons. After reviewing the properties of these solitons, we discuss several new aspects  
of the former. We solve the Bogomoln'yi equations 
perturbatively, to all orders in the inverse noncommutivity parameter, and show that the 
metric on the moduli space of $k$ vortices reduces to the computation of the trace of 
a $k\times k$-dimensional matrix. In the limit of large noncommutivity, we present an explicit 
expression for this metric.

\end{abstract}

\vskip 1 cm
\begin{center}
{\em Invited contribution to special issue of J.Math.Phys. on \\  
``Integrability, Topological Solitons and Beyond''} 
\end{center}

\end{titlepage}

\subsubsection*{\em Introduction and Results}

Vortices are enigmatic objects. Despite the apparent simplicity of the first 
order equations, no analytic expression for the solution has 
been found. Moreover, the metric on the moduli space, encoding the interactions of two 
or more vortices, remains unknown. This is in stark contrast to higher 
co-dimension solitons, such as monopoles and instantons, where seemingly more complicated 
equations readily yield results. 

Progress may be made in the limit of far separated vortices. By considering the 
leading order forces experienced by moving vortices, Manton 
and Speight determined the asymptotic form of the low-energy dynamics \cite{ms}. 
Their expression 
contains an unknown coefficient that characterizes the exponential return to vacuum 
of the Higgs field. Although a direct analytic computation of this coefficient 
appears difficult, a prediction has been given based on  
T-duality in string theory \cite{me}, and is in agreement with previous numerical results  
\cite{vega}. 

Another approach to understanding the dynamics is to deform the background 
space on which the vortices live. A cunning choice of deformation may ensure 
that the Bogomoln'yi equations become tractable. 
For example, it was discovered long ago that the tricky vortex 
equation is replaced by Liouville's equation when the background is taken to be 
hyperbolic space \cite{witt}. Strachan subsequently showed that this simplification is 
sufficient to allow an explicit calculation of the moduli space metric \cite{strachan}. 
More recently, Baptista and 
Manton considered the case of $k$ vortices interacting on a sphere of area $A\sim 4\pi k$ 
\cite{bm}. An analytic expression for the metric was given in the limit as the area of the 
sphere shrinks to a critical value, $A\rightarrow 4\pi k$. Curiously, in this limit, the 
vortex motion exhibits a symmetry enhancement, from the underlying $SU(2)$ symmetry 
of the sphere to $SU(k+1)$. The physics behind this enhancement remains somewhat 
puzzling. 

Here, we shall again deform the background space so that the dynamics of vortices becomes 
tractable. This time, we take space to be the flat, noncommutative plane. In two spatial 
dimensions, noncommutivity is rather natural since  it breaks only the discrete parity 
symmetry, 
leaving the continuous rotational symmetry intact. Solitons in noncommutative geometry have 
been extensively studied in recent times (see \cite{reviews} for reviews). In 
particular, aspects of vortices in the noncommutative abelian Higgs model have been discussed 
in \cite{tifr,kias,arg}. As we shall review, noncommutivity yields a one-parameter 
family of metrics on the vortex 
moduli space, depending on $\gamma$, a dimensionless combination of the gauge coupling 
constant $e^2$, the Higgs expectation value $v$ and the noncommutivity parameter $\theta$,
\be
\gamma = \theta e^2 v^2
\label{gamma}\ee
It was shown by Bak, Lee and Park \cite{kias} that solutions to the Bogomoln'yi equations 
exist only for $-1 \leq \gamma \leq +\infty$.  At the critical point $\gamma = -1$, 
the vortex solutions coincide with the fluxon solutions discovered in \cite{bakpoly}. 
Here, the moduli space of vortices is endowed with the flat, singular metric on ${\bf C}^k/S_k$, 
and the moduli space approximation breaks down \cite{kias}. For $\gamma<-1$, there are no 
further solutions to the first order vortex equations, but the non-BPS fluxon solutions survive 
as localized solitons carrying magnetic flux. In contrast, for $-1<\gamma < 0$, these 
non-BPS fluxon solutions are unstable to decay into the BPS vortices which have lower mass. 
For $\gamma>0$, only the BPS vortex solutions exist. 
This scenario, which was developed in \cite{kias}, is summarized in the figure below.

%%%%%%%%%%%%%%%%%%%%%%%%%%%%%%%%%%%%%
\newcommand{\onefigure}[2]{\begin{figure}[htbp]

         \caption{\small #2\label{#1}(#1)}
         \end{figure}}
\newcommand{\onefigurenocap}[1]{\begin{figure}[h]
         \begin{center}\leavevmode\epsfbox{#1.eps}\end{center}
         \end{figure}}
\renewcommand{\onefigure}[2]{\begin{figure}[htbp]
         \begin{center}\leavevmode\epsfbox{#1.eps}\end{center}
         \caption{\small #2\label{#1}}
         \end{figure}}
%%%%%%%%%%%%%%%%%%%%%%%%%%%%%%%%%%%%

\begin{figure}[htb]
\begin{center}
\epsfxsize=5in\leavevmode\epsfbox{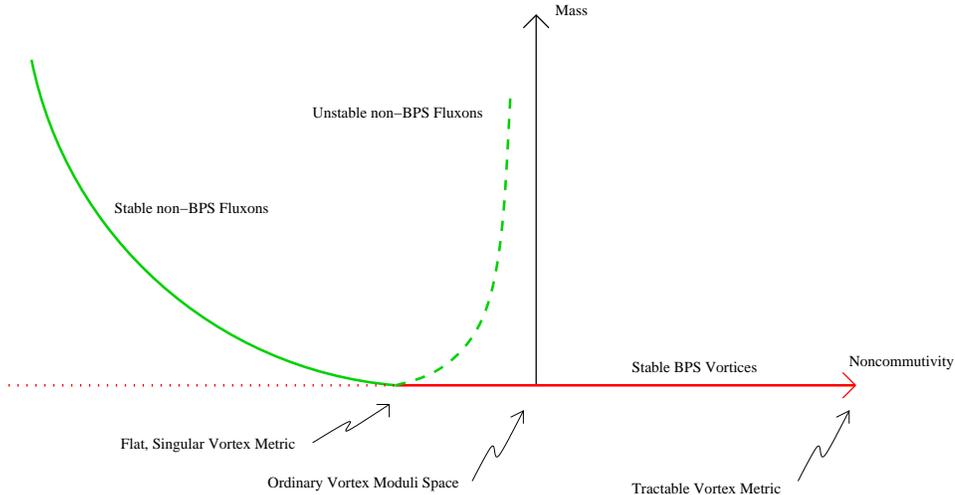}
\end{center}
\caption{The vortex phase diagram: BPS vortices exist for $-1\leq \gamma \leq \infty$, while 
non-BPS fluxons exist for $\gamma <0$, but are stable only for $\gamma\leq -1$.}
\end{figure}

Here we consider only $\gamma >0$, and present three, related, results. Firstly, we derive a 
$2k$-parameter formal solution to the vortex equations as an all-orders perturbative expansion 
in $\gamma^{-1}$. For arbitrary $\gamma$, we then show 
that the metric on the moduli space of $k$ vortices is given by the trace of a $k\times k$ matrix. 
This provides the noncommutative extension of Samols' expression in the ordinary commutative 
case \cite{samols}. Finally, in the limit $\gamma\rightarrow \infty$, we present an explicit 
expression for the 
metric on the moduli space of vortices. For $k$ vortices centered at $z_a\in {\mathbb C}$, 
$a=1,\ldots,k$, the K\"ahler potential for the multi-cover of the moduli space is given by
\be
{\cal K}=\log\det\exp(\bar{z}_az_b)
\label{k}\ee
Modding out by the permutation group $S_k$, exchanging the $z_a$, results in the true 
moduli space metric. 
This metric has appeared before in the study of noncommutative scalar solitons 
\cite{gms,lind1,ghs,lind2}, and we explain the similarities and differences with its 
appearance in the abelian Higgs model. 

The limit $\gamma\rightarrow\infty$ is customarily taken to mean large noncommutivity. 
However, as is clear from \eqn{gamma}, it may also be taken to be the strong coupling 
limit $e^2\rightarrow\infty$, which is commonly used in the context of gauged linear 
sigma models. Indeed, as we shall see, it is only in this limit that the proper 
kinetic energy of the noncommutative vortices remains finite. In the ordinary, commutative, 
abelian Higgs model, vortices become vanishingly small in this limit but, nevertheless, play an 
important role as singular worldsheet instantons \cite{witten}. We shall see that noncommutivity 
on the worldsheet resolves these singular vortices, in a manner similar to the resolution 
of singular $U(1)$ Yang-Mills instantons.

\subsubsection*{\em The Vortex Equations}

The Lagrangian of the abelian Higgs model at critical coupling is,
\be
{\cal L}=-\frac{1}{4e^2}F_{ij}F^{ij}+{\cal D}_i\phi {\cal D}^i\phi-\frac{e^2}{2}
(\phi\phi^\dagger-v^2)^2
\nn\ee
The model admits BPS vortices of mass $2\pi v^2k$, for any positive integer $k$, 
satisfying the first order equations of motion,
\be
F_{12}+e^2(\phi\phi^\dagger-v^2)=0\ \ \ ,\ \ \ {\cal D}_{\bar{z}}\phi=0\ \ \ ,\ \ \ 
\int F_{12}=2\pi k > 0
\label{vortwart}\ee
where we have introduced the complex structure $z=x_1+ix_2$ on the background space. 
Here we wish to consider the abelian Higgs model defined on the noncommutative complex plane, 
such that 
\be
[z,\bar{z}]=2\theta 
\nn\ee
It is common practice (see \cite{reviews}) to take $z$ to be an operator on the 
Hilbert space ${\cal H}$, isomorphic to the Hilbert space of a single harmonic 
oscillator. We define the usual 
creation and annihilation operators, satisfying $[a,a^\dagger]=1$, whose action 
on the orthonormal basis $|n\rangle =0,1,2,\ldots$ is given by,
\be
a|n\rangle = \sqrt{n}|n-1\rangle\ \ \ ,\ \ \ a^\dagger|n\rangle = 
\sqrt{n+1}|n+1\rangle
\nn\ee
For $\theta > 0$, the action of all spatial operators may thus be re-expressed as
\be
z=\sqrt{2\theta}a\ \ \ ,\ \ \ \partial_{z}\cdot=-\frac{1}{\sqrt{2\theta}}[a^\dagger,\cdot ]
\ \ \ ,\ \ \ \int d^2x=2\pi\theta {\rm Tr}
\nn\ee
where ${\rm Tr}$ is the trace over ${\cal H}$. Complex conjugation of the spacetime 
coordinate $z$ is identified with Hermitian conjugation of operators on ${\cal H}$. 
The fields $A_z$ and $\phi$ are 
themselves promoted to operators on ${\cal H}$. To simplify the equations, we 
rescale the Higgs field $\phi \rightarrow v\phi$, and decompose the gauge 
potential operator as
\be
A_{z}=\frac{i}{\sqrt{2\theta}}(a^\dagger + C^\dagger)
\nn\ee
The advantage of this notation is that the magnetic field $F_{12}$ is independent of $a$ and 
$a^\dagger$, and the vortex equations 
\eqn{vortwart} become the operator equations,
\be
1+[C^\dagger,C]&=&\gamma(\phi\phi^\dagger-1) \nn\\
\phi a + C\phi &=& 0 \label{vort}\\
{\rm Tr}(1+[C^\dagger,C]) &=& -k\ <\ 0
\nn\ee
As promised, the equations depend only on the dimensionless combination 
$\gamma=\theta e^2v^2$. They are valid only for $\gamma>0$.  
The rest of this paper will be devoted to analyzing these equations. 
However, for completeness, we firstly mention the extension to  
$\theta<0$. Defining $z=\sqrt{-2\theta}a^\dagger$, and 
$A_z=-i(a+C)/\sqrt{-2\theta}$, we have
\be
1+[C^\dagger, C]&=&\gamma(\phi\phi^\dagger - 1) \nn\\
\phi a^\dagger + C^\dagger\phi &=& 0 \label{antivortex}\\
{\rm Tr}(1+[C^\dagger,C])&=&+k >\ 0
\nn\ee
As depicted in Figure 1, while the equations \eqn{vort} are thought to have solutions 
for all $\gamma >0$ \cite{tifr}, 
equations \eqn{antivortex} have solutions only for $-1\leq \gamma \leq 0$ 
\cite{kias,arg}. 

Note that under a CP transformation, which is a symmetry of the theory, 
vortices are mapped to anti-vortices, while $\theta\rightarrow -\theta$. The phase 
diagram for anti-vortices is therefore given by reflecting Figure 1 in the vertical 
axis. The theory admits both BPS vortex and BPS anti-vortex solutions only 
for $|\gamma|\leq 1$. 

\subsubsection*{\em The Solution}

We turn now to the solution of the vortex equations \eqn{vort}. Perturbative progress 
can be made when the dimensionless parameter $\gamma$ is large. For 
coincident vortices, the solution to first order in $1/\gamma$, was given in \cite{tifr}. Here 
we present an iterative solution, for arbitrary vortex positions, to all orders in $1/\gamma$. 

Let us firstly consider the limit $\gamma\rightarrow\infty$. The first equation in \eqn{vort} 
now simply becomes $\phi_0\phi_0^\dagger = 1$, which can be partially inverted to give
\be
\phi_0^\dagger\phi_0=(1-P)
\nn\ee
where $P$, a projection operator on ${\cal H}$, determines the kernel of $\phi_0$: 
$Ker(\phi_0)=P{\cal H}$. In this limit, the gauge potential is given by 
$C_0=-\phi_0a\phi_0^\dagger$. After some manipulation, the second and 
third\footnote{In analyzing this final equation, it 
appears necessary to employ a suitable regularization 
of the trace over ${\cal H}$. We choose $Tr_N\,\cdot \equiv \sum_{n=1}^N\langle n|\cdot|n\rangle$, 
subsequently taking the limit $N\rightarrow\infty$.} 
vortex equations \eqn{vort} give further constraints on $P$,
\be
(1-P)aP&=&0\ \ \ \ ,\ \ \ \ {\rm Tr}P=k
\label{proj}\ee
Thus the kernel of $\phi_0$ is constrained to be a $k$-dimensional 
eigenspace of the annihilation 
operator $a$. Eigenvectors of $a$ are provided by the coherent states, 
\be
|z\rangle = \exp(za^\dagger)|0\rangle=\sum_{n=0}^\infty\frac{z^n}{\sqrt{n!}}|n\rangle
\nn\ee
which satisfy $a|z\rangle=z|z\rangle$ for any $z\in{\mathbb C}$. Thus, in general, 
solutions to the vortex equations in the limit $\gamma\rightarrow\infty$ are 
parameterized by $k$ complex vectors $|z_a\rangle$, $a=1,\ldots,k$ spanning 
$P{\cal H}$, with
\be
P=\sum_{a,b=1}^k|z_a>(h^{-1})^{ab}\langle z_b|
\label{p}\ee
where the overlap matrix $h$ is defined by
\be
h_{ab}=
\langle z_a|z_b\rangle=\exp(\bar{z}_az_b)
\label{h}\ee
We therefore find a $2k$ (real) parameter family of solutions. Hearteningly, 
the number of moduli is in agreement with the ordinary, commutative, vortex equations 
\cite{erick}. In the limit of far-separated vortices, we may think of the $z_a$ 
as the positions of $k$ unit-flux solitons. Note that this 
description of $P{\cal H}$ becomes singular in the 
limit as $|z_a\rangle\rightarrow |z_b\rangle$ for $a\neq b$. However, as 
explained nicely in \cite{ghs}, the underlying eigen-subspace remains smooth in this 
limit, and is spanned by $|z_a\rangle$ and $a^\dagger|z_a\rangle$.

To extend this analysis away from the $\gamma\rightarrow\infty$ limit, we make 
the expansion,
\be
\phi=\sum_{m=0}^\infty\gamma^{-m}\phi_m\ \ \ \ ,\ \ \ \ C=\sum_{m=0}^\infty\gamma^{-m}C_m
\nn\ee
The resulting iterative equations do not immediately determine the action of 
$\phi_1$ on the eigenspace $P{\cal H}$. To resolve this, we make 
the ansatz that the kernel of $\phi$ is independent of $\gamma$. In other words,
\be
Ker(\phi) = P{\cal H}
\nn\ee
I have not been able to derive this explicitly from \eqn{vort}, but have been unable 
to find solutions in which it is not the case. Proceeding with this assumption, 
we may express the solution to \eqn{vort} as
\be
C_m=-\phi_0d_m\phi_0^\dagger\ \ \ \ ,\ \ \ \ \phi_m=\phi_0\psi_m(1-P)
\nn\ee
where the operator coefficients $\psi_m$ and $d_m$ are determined by induction,  
starting from $\psi_0=1$, and with
\be
d_m&=&\psi_ma+\sum_{l=0}^{m-1}d_l(1-P)\psi_{m-l}\nn\\
\psi_m&=&\ft12(1-P)\sum_{l=0}^{m-1}
\left( d_l^\dagger (1-P)d_{m-l-1}-d_{m-l-1}(1-P)d_l^\dagger\right) 
-\ft12\sum_{l=1}^{m-1}\psi_l\psi_{m-l}^\dagger
\label{sol}\ee
To summarize, this perturbative solution is uniquely determined by a choice of 
projector \eqn{p} from among the $2k$ parameter family of suitable projectors. 
An important, open, problem is to determine the radius of convergence of 
this expansion.

\subsubsection*{\em The Low-Energy Dynamics}

Let us turn now to the low-energy dynamics of the noncommutative vortices. As usual, 
we consider the moduli space approximation, in which only the collective coordinates 
$z_a$, which determine the projection operator $P$, are allowed to vary in time. 
The linearized Bogomoln'yi equations \eqn{vort} are,
\be
[\dot{C}^\dagger,C]+[C^\dagger,\dot{C}]&=&\gamma(\dot{\phi}
\phi^\dagger+\phi\dot{\phi}^\dagger) \nn\\
\dot{\phi}a+\dot{C}\phi+C\dot{\phi}&=&0
\label{lin1}\ee
and are to be augmented with Gauss' law, the equation of motion for $A_0$. 
In our operator notation, this reads
\be
-[\dot{C}^\dagger,C]+[C^\dagger,\dot{C}]&=&\gamma(\dot{\phi}
\phi^\dagger-\phi\dot{\phi}^\dagger) 
\label{lin2}\ee
which can therefore be combined with the first of the Bogomoln'yi equations to 
give,
\be
[C^\dagger,\dot{C}]=\gamma\dot{\phi}\phi^\dagger
\nn\ee
The low-energy dynamics of the solitons is inherited from the kinetic energy terms 
of the original field theory in the standard Manton manner,
\be
T=2\pi\theta v^2 \ {\rm Tr}\left(\frac{1}{\gamma}\dot{C}^\dagger\dot{C}+
\dot{\phi}^\dagger\phi\right) \equiv 2\pi\theta v^2\ g_{ab}(z)\dot{z}^a\dot{\bar{z}}{}^b
\label{led}\ee
In the second equality above, we have anticipated the K\"ahlerity of the metric. This property 
is guaranteed by supersymmetry. To see this, a standard trick is to embed the theory in 
one with maximal supersymmetry, living in the maximal spacetime dimension. The 
abelian Higgs model may be embedded in a $d=5+1$ dimensional theory which is 
endowed with ${\cal N}=1$ supersymmetry (or 8 supercharges). The vortex solutions under 
consideration now become BPS 3-branes which 
preserve half of the supersymmetry. The addition of noncommutivity in two, transverse, 
spatial directions does not alter this fact, and the low-energy dynamics is thus described 
by a $d=3+1$, ${\cal N}=1$ (4 supercharges) non-linear sigma-model with target space 
given by the noncommutative vortex moduli space. The metric on the target space is 
necessarily K\"ahler. 

We start our analysis of the moduli space metric by once again taking the limit 
$\gamma \rightarrow \infty$. As is clear from \eqn{led}, the low-energy Lagrangian 
remains finite if we interpret this as $e^2\rightarrow\infty$. 
Since the kinetic terms for the gauge field become negligible 
in this limit, we have simply
\be
\frac{T_0}{2\pi\theta v^2}={\rm Tr}\ \dot{\phi}_0^\dagger\dot{\phi}_0
\nn\ee
From \eqn{lin1} and \eqn{lin2} we find that $\dot{\phi}_0=\dot{\phi}_0P=-\phi_0\dot{P}$, 
from which we derive the low-energy dynamics purely in terms of the projection operator 
$P$,
\be
\frac{T_0}{2\pi\theta v^2}={\rm Tr}\ P\,\dot{\phi}_0^\dagger\dot{\phi}_0=
\frac{1}{2} {\rm Tr}\ \dot{P}\dot{P}
\label{pdot}\ee
From the definition of the projection operator \eqn{p} in terms of the overlap 
matrix \eqn{h} it is simple to derive the explicit form of the metric,
\be
\frac{T_0}{2\pi\theta v^2}=  {\rm tr}\,( \partial_a \bar{\partial}_b \log h )\,
\dot{z}^a \dot{{\bar{z}}}{}^b
\label{zdot}\ee
where ${\rm tr}$ denotes the trace over the $k\times k$ matrix indices of $h$. 
The K\"ahler potential is therefore given by the expression \eqn{k} as promised. The 
expressions \eqn{pdot} and \eqn{zdot} have appeared before in the context of 
noncommutative  
solitons \cite{gms,lind1,ghs,lind2}. Let us pause briefly to review that work and 
explain the differences with the present case. The seminal work \cite{gms} considered 
pure scalar field theories in noncommutative spacetimes. It was shown that, in the 
limit $\theta\rightarrow\infty$, {\it any} projection operator solves the 
equation of motion. To proceed to finite theta one may work, as we have above, 
perturbatively in $1/\tilde{\gamma}=1/m^2\theta$ where $m$ is some mass scale of the theory. 
It was shown that, at first order in $1/\tilde{\gamma}$, only some projection operators 
survive as solutions to the equations of motion \cite{ghs,lind2}. These are precisely 
those operators satisfying \eqn{proj} above. At next-to-leading order in 
$1/\tilde{\gamma}$, these projectors too are lifted, and only isolated solutions 
remain. This scenario left certain aesthetic puzzles. For example, it was unclear why, a 
priori, the moduli space need be K\"ahler since the original field theory could not 
be embedded in a supersymmetric context and the solitons were not BPS. Moreover, the 
relevant solitons were not the most general solutions to any equations of motion, but 
rather the surviving approximate solutions at first order in perturbation theory.

In contrast, the appearance of the moduli space in the current context is more 
natural. It now appears at {\it zeroth} order in perturbation theory, rather 
than first, and the solitons are therefore solutions to certain equations of motion 
(namely those derived in the strict $\gamma\rightarrow\infty$ limit). Furthermore, 
as explained above, the K\"ahler nature of the target space finds an explanation 
in terms of supersymmetry.

The explicit metric for noncommutative vortices may be easily extracted from 
\eqn{zdot}. For $k\geq 3$, the algebra becomes somewhat entangled, but for 
$k=2$ it is simple and was given previously in \cite{lind1,ghs,lind2}. Factoring 
off an overall center of mass, we have the relative moduli space described in terms 
of the separation $z=z^1-z^2$ with metric,
\be
ds^2=\left(\frac{1}{2}\coth(|z|^2/2)-\frac{|z|^2}{4\sinh^2(|z|^2/2)}\right) dzd\bar{z}
\nn\ee
Since the two vortices are indistinguishable, we should orbifold this space by the $Z_2$ 
action $z\rightarrow -z$. It is simple to see that this renders the metric non-singular 
at the origin $z=0$.

We would like to extend this discussion beyond the $\gamma\rightarrow\infty$ limit. 
One may consider proceeding by calculating the contributions to the low-energy dynamics 
perturbatively in $1/\gamma$. Unlike the case of scalar field theories, where these effects 
induce a potential on the moduli space \cite{ghs,lind2}, for the case of vortices they merely 
correct the metric on the moduli space. The leading order contribution is given by 
$T=T_0+T_1/\gamma$ where
\be
\frac{T_1}{2\pi\theta v^2}={\rm Tr}\ P\left( \dot{P}\dot{P}a(1-P)a^\dagger 
-a(1-P)\dot{P}\dot{P}a^\dagger \right)
\nn\ee
However, this perturbative path appears tedious and illuminates little. 

Instead, we finish by 
deriving the noncommutative extension of Samols' localization theorem \cite{samols}. 
Recall that Samols analyzed the dynamics of $k$ vortices in 
the ordinary, commutative, abelian-Higgs model. Upon integrating the usual 
overlap of zero modes over the complex plane, he found that all contributions 
vanish apart from those arising at the $k$ zeroes of the Higgs field.

Examining our expressions for $T_0$ and $T_1$ above, we see that a similar 
phenomenon has occurred. The trace over the infinite dimensional 
Hilbert space ${\cal H}$ has been reduced to a more manageable trace over a 
$k$-dimensional subspace $P{\cal H}$. Here we show that this property holds for all 
values of $\gamma$. In order to derive this result, I have found it necessary to introduce the 
operator $\phi^{-1}$. Since $Ker(\phi)\neq 0$, we must define this operator with 
care. We require, 
\be
\phi\phi^{-1}=1\ \ \ ,\ \ \ \phi^{-1}\phi=1-P
\nn\ee
While the existence of such an operator in not guaranteed for all $\gamma$, it is 
a simple matter to construct it explicitly within the perturbative context of the solution 
\eqn{sol}, 
\be
\phi^{-1}=\sum_{m=1}^\infty\gamma^{-m}\tilde{\psi}_m\phi_0^\dagger
\nn\ee
where the operators $\tilde{\psi}_m$ are defined iteratively as $\tilde{\psi}_0=1$ and 
$\tilde{\psi}_1=-\psi_1$ with the remainder given by 
$\tilde{\psi}_m=-\sum_{l=1}^m\psi_l\tilde{\psi}_{m-l}$. 
The result below is therefore only strictly valid for $\gamma$ within the radius of 
convergence of the series \eqn{sol}.
Wielding this operator 
allows us to invert the Bogomoln'yi equation to $C=-\phi a \phi^{-1}$, supplying the 
leverage necessary to prise open the expression for the kinetic energy \eqn{led}. A little 
algebra reveals the final result,
\be
T=2\pi\theta v^2\ {\rm Tr}\ P\left(\dot{\phi}^\dagger\dot\phi-\frac{1}{\gamma}
a{\phi}^{-1}\dot{C}^\dagger\dot{\phi}\right)
\nn\ee
which indeed reduces to the trace  over the $k$-dimensional subspace $P{\cal H}$ as 
advertised.

\subsubsection*{\em Acknowledgments}

Thanks to Jan Troost and Ashwin Vishwanath for stimulating discussions. I'm supported by a 
Pappalardo fellowship, and thank  Neil Pappalardo for the money. This work was also 
supported in part by funds provided by the U.S. Department of Energy (D.O.E.) under 
cooperative research agreement \#DF-FC02-94ER40818.

\end{document}